# Bianisotropics and electromagnetics

E.O. Kamenetskii

Department of Electrical and Computer Engineering, Ben Gurion University of the Negev, Beer Sheva, 84105, Israel

January 19, 2006

**Abstract**

Bianisotropics is conceived as a physical concept describing electromagnetic media which possess intrinsic mechanisms of magnetoelectric coupling. We propose the main idea of bianisotropics as a combination of two physical notions: (a) the near-field manipulation and (b) chirality. From classical laws these two notions are in an evident contradiction since manifestation of chirality in electromagnetics is usually due to effects of the field non-locality. In attempts to combine the field locality and chirality in media properties, we come to non-classical aspects, which constitute a subject of a macroscopic quantum mechanics analysis. This opens a new field of studies, which we call as the problem of bianisotropic metamaterials (BMMs).

# 1. Introduction

Electromagnetics is a branch of a physical science that deals with the physical relations between electricity and magnetism. Bianisotropics concerns the subject of an intrinsic magnetoelectric (ME) coupling in media. The concept of a bianisotropic medium was coined in 1968 by Cheng and Kong [1] defining a medium with the most general linear constitutive relations. It was supposed that such generalized-form constitutive relations may unify two separate branches of research on moving media and magnetoelectric crystals.

On microwaves, bianisotropic media are conceived as artificial structures. The fact that small metallic resonant particles may show certain ME-like effects is well-known. It constitutes the subject of numerous theoretical and, not so numerous, experimental papers on microwave composite chiral and bianisotropic materials based on a collection of special-form metallic particles (see e.g., [2]). It is supposed that since particulate composite media are synthesized by artificially fabricated inclusions in a host medium, the designer is provided with large collections of necessary parameters: induced electric and magnetic moments with cross-polarization coupling. In an assumption that the inclusion sizes and spacings are small compared to the wavelength, a typical formal way of derivation of the macroscopic constitutive relations for bianisotropic composites is based on the quasistatic theories of polarization (see e.g. [3]).

It should be clear, however, that any quasistatic theories (similar, for example, to the quasistatic Lorentz theory used for artificial dielectrics [4]) are not applicable for metallic-inclusion bianisotropic composites. Actually, in a classical situation the "ME coupling" in composite media appears due to effects of non-local electromagnetic scattering from inclusions, but not because of the near-field cross-polarization effects. As it is discussed in [5], for interaction of the wave with such a "ME object" not just only the fields but also the various field gradients should play a role. It is stated in [6] that the separation between the macroscopic and microscopic electromagnetic descriptions is not quite as sharp in known artificial bianisotropic media as it is in pure dielectrics due to the fact that the cross-polarization coupling vanishes in the long-wavelength limit.

If one supposes that he has created an "artificial atom" with the local cross-polarization effect one, certainly, should demonstrate a special ME field in the near-field region. It means that using a gedankenexperiment with two quasistatic, electric and magnetic, point probes for the ME near-field characterization, one should observe not only an electrostatic-potential distribution (because of the electric polarization) and not only a magnetostatic-potential distribution (because of the magnetic polarization), one also should observe a special cross-potential term (because of the cross-polarization effect). This fact contradicts to classical electrodynamics. One cannot consider (classical electrodynamically) two coupled electric and magnetic dipoles – the ME particles – as local sources of the electromagnetic field [7]. So in a presupposition that an "artificial atom" with the near-field cross-polarization effect is really created, one has to show that in this particle there are special internal dynamical motion processes different from the classical motion processes.

The cross-polarization matrices in constitutive relations of bianisotropic media are, in fact, pseudotensors. This presumes symmetry breakdown in a structure of a medium. In different physical problems ME coupling is due to symmetry breaking phenomena. In crystals and molecular systems, magnetoelectricity – a phenomenon with local coexistence of electric and magnetic moments – takes place when space inversion is locally broken [8]. Following ideas of some recent theories, one sees that magnetoelectric interactions in crystal structures arise from toroidal distributions of currents and are described by so-called anapole moments [9]. The anapole moment takes place in systems with the parity violation and with the annual magnetic field [10]. At present, the role of anapole moments is considered as an important factor in understanding chirality (helicity) in different atomic, molecular and condense-matter phenomena. The anapole moment plays the essential role in nuclear helimagnetism [11,12]. It was considered as an intrinsic property of a diatomic polar molecule [13].

Usually, the effects of magnetoelectricity in natural crystals are considered without any relations with a symmetry structure of the EM field. On the other hand, at present, we can see a strong interest in effects of helicity in different free-space electromagnetic processes (both optical and microwave). These effects, realizing due to spiraling Poynting vector demonstrate, for example, the angular moments of light beams [14] and vortex-beam structures [15]. There are, in fact, the far-field phenomena of helicity, which are absolutely not related to the near-field manipulation.

Putting forth a physical concept of an electromagnetic composite material which possesses an intrinsic mechanism of local magnetoelectric coupling, we propose the main idea of bianisotropics as a combination of two physical notions: (a) the near-field manipulation and (b) chirality. From classical laws these two notions are in an evident contradiction since manifestation of chirality in electromagnetics is usually due to effects of the field non-locality. This leads one to come to conclusion that the unified ME fields originated by local ME particles should appear (in the near-field region) with the symmetry properties distinguishing from that of the electromagnetic fields. Most likely, there should be the symmetry properties demonstrated in the effects of natural magnetoelectricity and chirality known in atomic and molecular physics and in modern elementary particle physics. Following Fryberger [16], we will pursue the statement that the local "junction of electricity and magnetism should be at the source".

## 2. Metamaterials, bianisotropics and near-field structures

Maxwell's equations are largely useless until the relations between the quantities $\vec{D}, \vec{B}, \vec{E}$ and $\vec{H}$, which appear in them, have been established. This fact determines necessity of so-called constitutive relations for material parameters and leads to appearance of a special discipline – electrodynamics of continuous media [17]. The question, however, arises: What one implies when he uses the words "continuous media", or "materials", in electrodynamics. This question acquires a special meaning in a view of a strong interest in recent propositions and intensive studies of different artificial electromagnetic (EM) structures called as "metamaterials", "left-handed materials", "chiral materials", "bianisotropic materials", etc.

The term "metamaterials" means literally "more comprehensive materials". It was supposed that by this term one stresses on the fact that new artificially created EM structures are able to show material properties unknown in nature. Recently proposed artificial media with negative electric permittivity and magnetic permeability parameters – the left-handed materials – demonstrate the "metamaterial properties". On the other hand, bianisotropic materials were conceptualized as generalized media which are characterized by the ME (or bianisotropic) effect. Since there are no natural analogs of such media in microwaves, one supposes that realization of bianisotropic materials could be possible as artificial structures. In this sense, artificial microwave bianisotropic materials also demonstrate the "metamaterial properties".

The above definitions leave, however, the questions still open. Should one consider photonic (electromagnetic) bandgap structures (which properties are derived from diffraction of radiation on the same scale as the EM-wave wavelength) as metamaterials or there should be local media where the "particles" interact strongly with the near fields? Should be there electric or magnetic dipole-carrying excitation (plasmon oscillations, for example) in structure elements to get the metamaterial properties or not? In a view of increasing amount of new metamaterial propositions one, certainly, becomes faced with confusion in terminology. In attempts to arrange the question, we introduce here the following definitions.

We will call any construction of metal (perfect conductor) and/or lossless dielectric and/or lossless magnetic elements as a *diffraction-structure metamaterial* (DSM), when no "microscopic" motion equations ("microscopic" currents) are taken into account. There are metamaterials which gain the electromagnetic properties from their geometry rather than directly from microscopic physics of the natural materials they are composed of. So to describe properties of DSM only Maxwell equations have to be used. One can distinguish three types of DSMs. There can be near-field (or quasistatic) structures such as artificial dielectrics [4] or artificial magnets [18]. For near-field DSMs on uses the effective-medium parameters: the effective permittivity and permeability. Other examples of DSMs are non-local (field-gradient) structures. Microwave chiral and bianisotropic composite structures, based on special-form metallic particles, fall under this type of a DSM. In this case one can also use the effective-medium parameters for dilute composites. For such dilute composites one has the effective permittivity, permeability and effective parameters of ME coupling (see e.g. [2]). Photonic (electromagnetic) bandgap structures demonstrate the third type of DSMs. The properties of these DSMs are derived from diffraction of radiation and therefore are structured on the same scale as the EM-wave wavelength. There is the retardation-type interaction contrary to the instantaneous-type interaction in the near-field DSMs.

Together with DSMs we distinguish the *motion-equation metamaterials* (MEMs). The MEMs gain the electromagnetic properties mainly from "microscopic" motion equations for particles (quasiparticles) inside the natural materials they are composed of. There are, for example, electric and/or magnetic dipole-carrying excitations such as plasmon, exciton, and magnon oscillations. A wire-grid structure demonstrates an example of a MEM with plasmon oscillations [19, 20]. Another example of a plasmon-oscillation MEM is shown in [21, 22] where the dielectric properties of nanocomposites have been studied. The effective-medium permittivities of such MEMs appear from plasmon modes describing by the classical motion equations. There is a possibility to combine properties of DSMs and MEMs. These combinations may lead to appearance of very unique material structures. Combination of a near-field DSM [18] and a wire-grid MEM [19, 20] gives an example of well-known left-handed metamaterials [23]. Should one restrict consideration of MEMs only with classical motion equations? Certainly, when one extends an analysis for quantum (or quantum-like) motion equations, fundamentally new material properties can be expected. Recently, intensive studies of "artificial solids" based on a composition of quantum-dot "artificial atoms" have been launched [24]. Following our classification, such "artificial solids" fall under definition of MEMs. For MEMs with quantum (or quantum-like) motion equations, the following fundamental principle takes place: the structure is conceived as a local medium where "particles" interact with the near fields based on the Hamiltonian mechanism of interaction.

The above classification of metamaterials will help us to understand how the same idea of bianisotropics can be correlated with the material concept. One of the main aspects attracted the concept of metamaterials was a possibility for the near-field manipulation [25]. In such a sense, metamaterials can be characterized as structures with tailored electromagnetic response. For bianisotropics, understanding of physics of the near field plays a very important role. Usually, in electrodynamics the following classification of the EM fields is used: the far-field EM fields are considered as the propagating waves and the near-field EM fields – as the evanescent (exponentially decaying) modes. The importance of phenomena involving evanescent electromagnetic waves has been recognized over the last years. The fact that evanescent waves are more confined than the single tone sinusoid waves and hence contain wider range of spatial frequencies indicates that it may be possible to have no theoretical limit of resolution for the near-field patterns. At present, the near-field manipulation becomes an important factor in new applications, such as near-field microscopy and new material structures. From a general point of view, the near-field can be defined as the extension outside a given structure (sample) of the field existing inside this structure (sample). Physically, there can be distinguished different categories of the near-fields [26]. For our purpose, we will consider three types of the near-field EM structures.

A. Evanescent modes

For the Helmhotz equation

$$\varepsilon\mu\, k_0^2 = k_x^2 + k_y^2 + k_z^2,\tag{1}$$

where

$$k_0^2 = \frac{\omega^2}{c^2},\tag{2}$$

$k_x, k_y, k_z$ are wavenumbers along *x,y,z* axes in a medium, two solutions are possible when $k_x$ and $k_y$ are real quantities. The first solution corresponds to the case

$$\varepsilon\mu\, k_0^2 > k_x^2 + k_y^2.\tag{3}$$

It shows that $k_z$ is a real quantity and one has, as a result, the 3D propagating EM process. The second solution takes place when

$$\varepsilon\mu\, k_0^2 < k_x^2 + k_y^2.\tag{4}$$

So $k_z$ is an imaginary quantity. One has the 2D propagating EM process along *x* and *y* axis and the evanescent-mode fields (the near fields) in *z* direction. The typical examples of such evanescent modes can be demonstrated in the field structures of closed microwave waveguides and open optical waveguides. The importance of evanescent electromagnetic waves was recently realized in connection with emergence of near-field optics microscopy. Taking evanescent waves into account prevents the use of any approximations and requires the detailed solution of the full set of Maxwell equations [27]. As an effective method to study evanescent electromagnetic waves, for solving Maxwell's equations one can use expansion in multipoles where electromagnetic fields are constructed from scalar-wave eigenfunctions of the Helmholtz equation [27, 28].

It is evident that to get extremely big quantities of imaginary $k_z$ one should have extremely big quantities of real $k_x$ and $k_y$. On the way to create a perfect lens based on left-handed metamaterials [25], this fact may become the stronger limitation. Really, the concept of effective medium cannot be used for a perfect-lens slab of a left-handed metamaterial illuminated by a point source, when evanescent waves have a transverse wavelength of the order of or less than the dimensions of the inclusions or their spacings [29]. Misunderstanding of such a limitation can lead to serious flaws in physics of new material propositions for perfect-lens slabs. As an example, we can refer to paper [30], where a material for a perfect-lens slab is conceived as the dilute mixture of helical inclusions. The fact that inclusions are not in each other near field cast doubts on the vital effect of evanescent waves in such a slab lens.

B. Quasistatic limit

The quasistatic limit means $|\vec{k}_0| \to 0$ and so $|k_x|, |k_y|, |k_z| \to 0$ as well. In this case, no-wave time-dependable quasistatic fields exist. Such quasistatic electromagnetic fields can be realized only due to local sources: or local-capacitance alternative electric charges with surrounding potential electric fields:

$$\vec{E} = -\nabla\varphi,\tag{5}$$

or local-loop conductive electric currents with surrounding potential magnetic fields:

$$\vec{H} = -\nabla \psi. \tag{6}$$

Spatial distributions of potential $\varphi$ as well as potential $\psi$ are described by the Laplace equation. Examples are the quasistatic fields surrounding tip-structure probes in modern microwave-microscopy devices [31]. Certainly, there is no physical mechanism for possible ME coupling between such local electric and magnetic sources.

C. Quasistatic oscillations

The symmetry between the electric and magnetic fields is broken in finite temporally dispersive media. In this case, quasistatic oscillations may take place. For quasistatic oscillations

$$k_0 \ll 1/l, \tag{7}$$

where $l$ is the characteristic size of a body. In such oscillations, there are no electromagnetic retardation effects since one neglects or electric or magnetic displacement currents. The following are some examples of quasistatic oscillations.

(a) Quasistationary EM fields in small metal samples. These fields are described by Maxwell equations with neglect of the electric displacement currents. Inside a sample we have the "heat-conductivity-like" equation for the magnetic field:

$$\nabla^2 \vec{H} = \frac{4\pi\mu\sigma}{c^2} \frac{\partial \vec{H}}{\partial t}. \tag{8}$$

Outside a sample there are the quasistatic-field equations:

$$\nabla \cdot \vec{B} = 0, \quad \nabla \times \vec{H} = 0. \tag{9}$$

The solutions correspond to imaginary numbers of $k_x, k_y,$ and $k_z$ showing that there are non-stationary decaying fields [17].

(b) Plasmon-oscillation fields are the fields due to collective oscillations of electron density. When one considers a metal or a semiconductor as a composite of positive ions forming a regular lattice and conduction electrons which move freely through this ionic lattice, there can be longitudinal oscillations of the electronic gas – the plasma oscillations. The interface between such a sample and a dielectric may also support charge density oscillations – surface plasmons. In the case of surface plasmon modes, the surface plasmon field decays exponentially away from the interface. For the non-retardated electrostatic description (one neglects the magnetic displacement current), an electric field is the quasielectrostatic field ($\vec{E} = -\nabla \varphi$). The plasmon oscillations may be characterized by *electrostatic wave functions*, which are eigenfunctions of the Laplace-like equation. For negative frequency dependent permittivity, one can observe a discrete spectrum of propagating electrostatic modes [32] and electrostatic resonances [33]. Surface plasmons can interact with photons (with the same polarization state) if the momentum and energy conditions are right. There is a link between near-field focusing action and the existence of well-defined surface plasmons [25].

(c) Magnetostatic (MS) oscillations are observed in small temporally dispersive ferromagnet samples. For these quasistationary fields, non-retardated magnetostatic description (one neglects the electric displacement current) can be used. So a magnetic field is the quasimagnetostatic field: $\vec{H} = -\nabla \psi$. Inside a ferrite sample one has the Walker equation for *MS-potential wave function* $\psi$:

$$\nabla \cdot [\vec{\vec{\mu}}(\omega)\nabla \psi] = 0, \tag{10}$$

where $\vec{\vec{\mu}}$ is the permeability tensor. Outside a sample, there is the Laplace equation:

$$\nabla^2 \psi = 0. \tag{11}$$

For a negative diagonal component of the permeability tensor, the solutions inside a sample may be characterized by real wave numbers for all three dimensions. In this case one can observe a discrete spectrum of propagating magnetostatic modes and magnetostatic resonances [34].

There are no classical physics mechanisms for internal coupling between the electrostatic and magnetostatic oscillations. If one conceives realization of a sample with a simple combination of the plasmon and magnon sources, there will not be a local ME particle surrounded by the unified ME field. At the same time, the near-field structure of evanescent modes has a pure EM nature and, certainly, is not related to the local ME field. A bianisotropic metamaterial based on ME particles with the near-field cross-polarization effects (for such a near-field bianisotropic metamaterial we use an abbreviation BMM) cannot be realized using just only the Maxwell-equation analysis and should be conceived as a special MEM. To make this fact more clear we will dwell now on a formal structure of Maxwell equations in the context of bianisotropics.

## 3. Formal structure of Maxwell equations and bianisotropics

Usually, a bianisotropic medium is classified as the medium characterized by constitutive relations:

$$\vec{D} = \vec{\vec{\varepsilon}} \cdot \vec{E} + \vec{\vec{\xi}} \cdot \vec{H}, \tag{12}$$

$$\vec{B} = \vec{\vec{\zeta}} \cdot \vec{E} + \vec{\vec{\mu}} \cdot \vec{H}, \tag{13}$$

where $\vec{\vec{\varepsilon}}, \vec{\vec{\mu}}, \vec{\vec{\xi}},$ and $\vec{\vec{\zeta}}$ are the medium dyadics. From classical electrodynamics point of view, this medium is considered as the most general and exotic medium. It is supposed that studies of this medium provide deep insight into the nature of macroscopic Maxwell's equations, in general, and into the EM-field propagation mechanism, in particular. The results obtained by analytical means give the Green's dyadics, the field and source decomposition and show different aspects of the plane-wave propagation in bianisotropic media [35]. It is supposed that with the introduction of constitutive relations the axiomatic approach to classical electrodynamics is completed.

The question, however is: How one gets constitutive relations (12), (13)? One can postulate these relations, as a case of the most general medium, and just only see for himself if such relations do not violate the Maxwell equations. This way of proof really works and the main physical model is based on a simple assumption that the medium is a structure composed with particles of intrinsically coupled electric and magnetic dipoles. Certainly, in his book, Post writes [36]: "Macroscopically, there is no reason to assume that these [electric and magnetic] dipoles cannot be glued together, pairwise in an almost rigid manner". As it is stated by Gronwald *et al* [37], the constitutive relations for a general linear ME medium have to be just postulated as an axiom of classical electrodynamics. At the same time, it is pointed out [37] that derivation of these constitutive relations should be obtained after an averaging procedure from a microscopic model of matter. It is unclear, however, what kind of such a model is presumed by the authors.

As we all know (see e.g. [7, 17, 38]), electromagnetic fields in a medium arise from the microscopic Maxwell equations written for the microscopic electric $\vec{e}$ and magnetic $\vec{h}$ fields, microscopic electric charge density $\rho(\vec{r},t)$ and microscopic electric current density $\rho \vec{v}$:

$$\nabla \times \vec{e} = -\frac{1}{c}\frac{\partial \vec{h}}{\partial t}, \quad \nabla \cdot \vec{e} = 4\pi\rho, \tag{14}$$

$$\nabla \times \vec{h} = \frac{4\pi}{c}\rho \vec{v} + \frac{1}{c}\frac{\partial \vec{e}}{\partial t}, \quad \nabla \cdot \vec{h} = 0. \tag{15}$$

In natural media, the microscopic electric charge density and the microscopic electric current density are defined as:

$$\rho(\vec{r}) = \sum_{i,a} q_{i,a} \delta(\vec{r} - \vec{r}_a - \vec{\xi}_i), \quad \rho \vec{v} = \sum_{i,a} q_{i,a} \dot{\vec{\xi}}_i \delta(\vec{r} - \vec{r}_a - \vec{\xi}_i), \tag{16}$$

where $a$ is a number of a structural element (atom, molecule, etc.), $\vec{\xi}_i$ is a radius-vector of the $i^{th}$ electric charge counted from a center of the corresponding structural element, $\vec{r}_a$ is a radius-vector of the element center, $\dot{\vec{\xi}}_i = \frac{d\vec{\xi}_i}{dt}$. A theory of electromagnetic processes in media is called as macroscopic electrodynamics or electrodynamics of continuous media. This is a phenomenological theory. It makes sense for average quantities: average positions and velocities of particles, which constitute a medium, average electric and magnetic fields. The averaging procedure takes place in a *physically infinitesimal volume* – the volume whose sizes are much larger than the space between particles, but much smaller than the characteristic dimension of the macroscopic quantity variations. The macroscopic Maxwell equations and the continuity equation are the following:

$$\nabla \times \vec{E} = -\frac{1}{c}\frac{\partial \vec{B}}{\partial t}, \quad \nabla \cdot \vec{E} = 4\pi\langle\rho\rangle, \tag{17}$$

$$\nabla \times \vec{B} = \frac{4\pi}{c}\langle\rho \vec{v}\rangle + \frac{1}{c}\frac{\partial \vec{E}}{\partial t}, \quad \nabla \cdot \vec{B} = 0, \tag{18}$$

$$\frac{\partial \langle\rho\rangle}{\partial t} + \nabla \cdot \langle\rho \vec{v}\rangle = 0. \tag{19}$$

In these equations vector $\vec{E}$ is the electric field strength in a medium and means the averaged microscopic electric field strength. Vector $\vec{B}$ is the magnetic flux density in a medium, which means the averaged microscopic magnetic field strength. So

$$\langle\vec{e}\rangle = \vec{E} \quad \text{and} \quad \langle\vec{h}\rangle = \vec{B}. \tag{20}$$

We have also averaged electric charge density $\langle\rho\rangle$ and averaged electric current density $\langle\rho \vec{v}\rangle$.

For an electrically neutral and non-conductive medium we have

$$\langle \rho \vec{v} \rangle = \langle (\rho \vec{v})_{bound} \rangle = \frac{\partial \vec{P}}{\partial t} + c \nabla \times \vec{M}, \tag{21}$$

where

$$\vec{P}(\vec{r}) \equiv \left\langle \sum_{i,a} q_{i,a} \vec{\xi}_i \delta(\vec{r} - \vec{r}_a) \right\rangle \tag{22}$$

is an average density of an electric dipole moment and

$$\vec{M}(\vec{r}) \equiv \left\langle \sum_{i,a} \frac{q_{i,a}}{2c} (\vec{\xi}_i \times \dot{\vec{\xi}}_i) \, \delta(\vec{r} - \vec{r}_a) \right\rangle \tag{23}$$

is an average density of the magnetic moment in a medium. From the above equations one introduces the quantities:

$$\vec{D} \equiv \vec{E} + 4\pi \vec{P} \quad \text{and} \quad \vec{H} \equiv \vec{B} - 4\pi \vec{M} \tag{24}$$

and obtains, as a result, macroscopic Maxwell's equations for a non-conductive medium.

There is, however, another way to derive the macroscopic Maxwell's equations. Suppose formally that there exist both the electric and magnetic microscopic charges. For such a hypothetical case, one easily obtains the macroscopic Maxwell equations from the microscopic equations, but physical meaning of the averaged quantities is not the same as in the above case. The microscopic Maxwell equations are the following:

$$\nabla \times \vec{e} = -\frac{4\pi}{c} \rho^m \vec{v}^m - \frac{1}{c} \frac{\partial \vec{h}}{\partial t}, \qquad \nabla \cdot \vec{e} = 4\pi \rho^e, \tag{25}$$

$$\nabla \times \vec{h} = \frac{4\pi}{c} \rho^e \vec{v}^e + \frac{1}{c} \frac{\partial \vec{e}}{\partial t}, \qquad \nabla \cdot \vec{h} = 4\pi \rho^m \tag{26}$$

and the microscopic charge densities and the microscopic current densities are defined as:

$$\rho^e(\vec{r}) = \sum_{i,a} q_{i,a}^e \delta(\vec{r} - \vec{r}_a - \vec{\xi}_i), \qquad \rho^m(\vec{r}) = \sum_{j,b} q_{j,b}^m \delta(\vec{r} - \vec{r}_b - \vec{\xi}_j), \tag{27}$$

$$\rho^e \vec{v}^e = \sum_{i,a} q_{i,a}^e \dot{\vec{\xi}}_i^e \, \delta(\vec{r} - \vec{r}_a - \vec{\xi}_i), \qquad \rho^m \vec{v}^m = \sum_{j,b} q_{j,b}^m \dot{\vec{\xi}}_j^m \, \delta(\vec{r} - \vec{r}_b - \vec{\xi}_j). \tag{28}$$

Here $a$ and $b$ are numbers of electric and magnetic structural elements, respectively, $\xi_i^e$ is the radius-vector of the $i^{th}$ electric charge counted from the electric "atomic" center and $\xi_j^m$ is the radius-vector of the $j^{th}$ magnetic charge counted from the magnetic "atomic" center. For the averaged microscopic fields, we use the designations:

$$\langle \vec{e} \rangle = \vec{E} \quad \text{and} \quad \langle \vec{h} \rangle = \vec{H}. \tag{29}$$

Here vector $\vec{E}$ is the electric field strength in a medium which means the averaged microscopic electric field strength and vector $\vec{H}$ is the magnetic field strength in a medium which means the averaged microscopic magnetic field strength. The macroscopic Maxwell equations and the continuity equations are the following:

$$\nabla \times \vec{E} = -\frac{4\pi}{c} \langle \rho^m \vec{v}^m \rangle - \frac{1}{c} \frac{\partial \vec{H}}{\partial t}, \quad \nabla \cdot \vec{E} = 4\pi \langle \rho^e \rangle, \tag{30}$$

$$\nabla \times \vec{H} = \frac{4\pi}{c} \langle \rho^e \vec{v}^e \rangle + \frac{1}{c} \frac{\partial \vec{E}}{\partial t}, \quad \nabla \cdot \vec{H} = 4\pi \langle \rho^m \rangle, \tag{31}$$

$$\frac{\partial \langle \rho^e \rangle}{\partial t} + \nabla \cdot \langle \rho^e \vec{v}^e \rangle = 0, \quad \frac{\partial \langle \rho^m \rangle}{\partial t} + \nabla \cdot \langle \rho^m \vec{v}^m \rangle = 0. \tag{32}$$

We introduce average densities of electric and magnetic dipole moments as

$$\vec{P}^e(\vec{r}) \equiv \left\langle \sum_{i,a} q_{i,a} \vec{\xi}_i^e \delta(\vec{r} - \vec{r}_a) \right\rangle, \quad \vec{P}^m(\vec{r}) \equiv \left\langle \sum_{j,b} q_{j,b} \vec{\xi}_j^m \delta(\vec{r} - \vec{r}_b) \right\rangle. \tag{33}$$

For a non-conductive medium with dipolar particles we have for the average current densities

$$\langle \rho \vec{v} \rangle^e = \langle (\rho \vec{v})^e_{bound} \rangle = \frac{\partial \vec{P}^e}{\partial t}, \quad \langle \rho \vec{v} \rangle^m = \langle (\rho \vec{v})^m_{bound} \rangle = \frac{\partial \vec{P}^m}{\partial t}. \tag{34}$$

We put now the above equations in the macroscopic Maxwell equations:

$$\nabla \times \vec{E} = -\frac{4\pi}{c} \frac{\partial \vec{P}^m}{\partial t} - \frac{1}{c} \frac{\partial \vec{H}}{\partial t}, \quad \nabla \cdot \vec{E} = -4\pi \nabla \cdot \vec{P}^e, \tag{35}$$

$$\nabla \times \vec{H} = \frac{4\pi}{c} \frac{\partial \vec{P}^e}{\partial t} + \frac{1}{c} \frac{\partial \vec{E}}{\partial t}, \quad \nabla \cdot \vec{H} = -4\pi \nabla \cdot \vec{P}^m. \tag{36}$$

With introduction of two quantities:

$$\vec{D} \equiv \vec{E} + 4\pi \vec{P}^e \quad \text{and} \quad \vec{B} \equiv \vec{H} + 4\pi \vec{P}^m \tag{37}$$

one obtains, as a result, the standard-form macroscopic Maxwell's equations for a non-conductive medium.

The above analysis shows that formally equations of macroscopic electrodynamics are equally applicable for media with electric polarization and magnetization – the EP&MAG media – as well as for media with electric polarization and magnetic polarization – the EP&MP media. Of course, microscopic magnetic charges and microscopic magnetic currents are not known in modern electromagnetism [7]. Nevertheless, for artificial structures composed by local elements one can formally use such notions of "microscopic" magnetic charges and "microscopic" magnetic currents. Some known artificial material structures with local properties fall under the above classification of the EP&MAG and EP&MP media. In particular, the left-handed metamaterials described in [23] are the EP&MAG media. On the other hand, based on recent propositions and analysis of magnetic-dot arrays with magnetic dipolar interactions [39.40] one can conceive the microwave EP&MP media when such magnetic dipoles are mixed with small electric dipoles [21,22]. In such metamaterials, the dot near-fields are originated from "microscopic" electric and magnetic charges. This makes our consideration of microscopic Maxwell's equations with both types of charges not so hypothetical.

The motion equations for both above cases are local equations: the average procedure for microscopic current densities takes place in scales much less than a scale of variation of any macroscopic quantity. An important thing is that no magnetoelectric couplings on the microscopic level are assumed in these motion equations. To get the ME coupling for the EP&MAG media on the microscopic level one should suppose that there exist helical motion processes for electrons in media. Such a helical motion really may take place for free charged particles in cross, electric and magnetic, fields. But, in frames of a classical description, no helical loops (recursion motions) are possible for bound charges. For the EP&MP media one cannot envisage any possibility for the microscopic-level ME coupling as well, since no classical laws describe interaction between linear electric and magnetic currents. The symmetry properties of magnetic charge and current densities under both spatial inversion and time reversal are opposite to those of electric charge and current densities [7]. One may expect realizing local ME coupling only when dynamical symmetry breaking occurs. As it was pointed out by Fryberger [16], electromagnetism and magnetoelectricity have not yet been unified. There are no classical equations of motion in which the fields associated with electrically charge particles will exert forces on magnetically charge particles, and vice versa. It was discussed in [16] that the required terms describing cross interactions may appear in a natural way when generalized electromagnetism is formulating using Dirac algebra.

Fig. 1 shows a model of an EP&MP material. A model of a questionable EP&MP medium with local ME coupling is shown in Fig.2. As we will discuss below, there could be found a possible solution to realize such a medium based on certain quantum-like mesoscopic resonant particles.

Returning to the above classical analysis we state once again that both EP&MAG and EP&MP media do not possess the microscopic ME coupling. However, on the macroscopic level of consideration one can obtain the ME coupling in a frame of a classical description. So the Post statement that *macroscopically* there is no reason to assume that the electric and magnetic dipoles cannot be glued together [36], becomes evident. The averaged quantities of the charge density and current density appear from the microscopic quantities. On the other hand, these averaged quantities are functions of the electric and magnetic fields. Besides this, the current density could be a function of the time and space variations of the fields in a medium. Let us come back to macroscopic Maxwell equations for EP&MAG media. In a most general form we have [41]:

$$\langle \rho \vec{v} \rangle = f\left( \vec{E}, \vec{B}, \frac{\partial \vec{E}}{\partial t}, \frac{\partial \vec{B}}{\partial t}, \frac{\partial E_i}{\partial x_k}, \frac{\partial B_i}{\partial x_k} \right), \qquad (38)$$

where subscripts $i$ and $k$ denote the field components and space coordinates. For linear electromagnetic processes we should take into account linear terms of function $f$. The quantity $\langle \rho \vec{v} \rangle$ is a polar vector. So the terms of function *f* should not be scalars and axial vectors, but only polar vectors. In a general case of a homogeneous medium, we can write:

$$\langle \rho \vec{v} \rangle = \vec{\vec{\sigma}} \cdot \vec{E} + \vec{\vec{\kappa}} \cdot \frac{\partial \vec{E}}{\partial t} + \vec{\vec{\gamma}} \cdot \frac{\partial \vec{B}}{\partial t} + c\vec{\vec{\alpha}} \cdot \nabla \times \vec{B} + c\vec{\vec{\delta}} \cdot \nabla \times \vec{E}, \qquad (39)$$

where $\vec{\vec{\sigma}}, \vec{\vec{\kappa}}, \vec{\vec{\alpha}}$ are tensors and $\vec{\vec{\gamma}}, \vec{\vec{\delta}}$ are pseudotensors. These tensors and pseudotensors are dependable on the medium properties. The coefficient *c* (the light velocity) we introduced here arbitrarily. For an ordinary non-conductive medium one has $\vec{\vec{\sigma}} = \vec{\vec{\gamma}} = \vec{\vec{\delta}} = 0$ and Eq. (39) can be evidently reduced to Eq. (21).

We consider now a non-trivial case of a non-conductive medium. Since for EP&MAG there are $\vec{\vec{\sigma}} = 0$ and $\vec{\vec{\gamma}} = 0$ (there are no linear magnetic currents), one has from Eq. (39)

$$\langle \rho \vec{v} \rangle = \frac{\partial \vec{P}}{\partial t} + c \nabla \times \vec{M} + c\vec{\vec{\delta}} \cdot \nabla \times \vec{E}. \qquad (40)$$

The first two terms in a right-hand side of this equation are the same as the corresponding terms in Eq. (21). The third term in a right-hand side of Eq. (40) shows the non-locality effect. When one uses the Maxwell equation $\nabla \times \vec{E} = -\frac{1}{c}\frac{\partial \vec{B}}{\partial t}$, this term looks like the ME term.

The ME properties due to non-locality on the macroscopic level become especially evident when one considers the EP&MP media. Since $\langle \rho \vec{v} \rangle^e$ is a polar vector and $\langle \rho \vec{v} \rangle^m$ is an axial vector and taking into account that in EP&MP media there are only linear electric and magnetic currents, we can write as a result:

$$\langle \rho \vec{v} \rangle^e = \frac{\partial \vec{P}^e}{\partial t} + c\vec{\vec{\delta}} \cdot \nabla \times \vec{E} \qquad (41)$$

and

$$\langle \rho \vec{v} \rangle^m = \frac{\partial \vec{P}^m}{\partial t} + c\vec{\vec{v}} \cdot \nabla \times \vec{H}, \qquad (42)$$

where $\vec{\vec{\delta}}$ and $\vec{\vec{v}}$ are pseudotensors, which are dependent on the medium properties. The second terms in a right-hand side of these equations show the non-locality effect. Based on Eqs. (41), (42) and taking into account Maxwell's equations, one may obtain finally constitutive relations in a form of Eqs. (12), (13).

Following a general representation of electromagnetic properties of a medium, we showed that bianisotropic constitutive relations appear as the effect of non-locality (effect of field space derivatives) on the macroscopic level of consideration. There are no classical models on the microscopic level, which demonstrate the ME effect. For this reason, classical theories for homogenization of local bianisotropic composites based on the Maxwell Garnett and the Bruggeman formalism (see e.g. [42]), look as physically groundless. It is also relevant to mention here about some "bianisotropic effects in left-handed materials" discussed in recent publications. The electric and magnetic responses in the left-handed materials are supposed to be independent from each other [23]. Nevertheless, Marques *et al* stated in [43] that in split-ring-resonator (SRR) particles one should observe the ME effect. This statement found further consideration, both in theory and experiments [44, 45]. It follows, however, that the physics which underlies such effects is due to electromagnetic scattering from small non-local objects, but not because of the near-field (local) cross-polarization effects in ME particles. A SRR is an electric contour. It does not have a non-electromagnetic (motion-equation) discrete oscillating spectrum. So one cannot expect that the near-field region surrounding this particle will show special ME properties different from the known electromagnetic near fields. As we discussed above, one cannot consider (classical electrodynamically) two coupled electric and magnetic dipoles – the ME particles – as point sources of the electromagnetic field. So the "ME effect" in SRR observed in [43 - 45] and the actual microscopic ME effect bear no resemblance to one another. In a presupposition that an "artificial atom" with the cross polarization effect is really created, one has to show that in this particle there are special internal dynamical processes distinguishing from classical electromagnetic processes.

## 4. Field polarization structure in local bianisotropic media

Let us suppose, however, that local bianisotropic media exist. In such a formal representation (still without necessary justification of the microscopic ME effect and, therefore, without a well-justified microscopic theory of homogenization), unique properties of the field polarization structure in these media become evident even from a macroscopic classical analysis.

In electrodynamics of continuous media, the integral-form constitutive relations (ICRs) represent the most general form of constitutive relations [17, 46]. Based on such a representation, we can characterize a priori a bianisotropic medium – a generalized linear medium – as a structure described by four integral operators [6]. Formally, there can be two versions of the ICRs for a bianisotropic medium. The first version is

$$
\begin{aligned}
D_i(t,\vec{r}) &= (\alpha_{ij} \circ E_j) + (\beta_{ij} \circ B_j), \\
H_i(t,\vec{r}) &= (\gamma_{ij} \circ E_j) + (\nu_{ij} \circ B_j)
\end{aligned}
\tag{43}
$$

and the second version is

$$
\begin{aligned}
D_i(t,\vec{r}) &= (\varepsilon_{ij} \circ E_j) + (\xi_{ij} \circ H_j), \\
B_i(t,\vec{r}) &= (\zeta_{ij} \circ E_j) + (\mu_{ij} \circ H_j).
\end{aligned}
\tag{44}
$$

In the right-hand sides of these expressions we have the integral operators similar to the following integral operator:

$$
(\alpha_{ij} \circ E_j) = \int_{-\infty}^{t} dt' \int d\vec{r}' \, \alpha_{ij}(t,\vec{r},t',\vec{r}') E_j(t',\vec{r}') .
\tag{45}
$$

In the above equations, sub-subscripts $i$ and $j$ correspond to Cartesian coordinates.

The kernels of the operators in the above ICRs are "responses" of a medium to the $\delta$-function electric and magnetic fields. So the convergence of integrals in the ICRs can be proven if one shows a physical mechanism of influence of short-time and short-space electrostatic and magnetostatic interactions on the medium polarization properties. In such a case the above ICRs describe a bianisotropic media with local properties. The ICRs (43) correspond to the case of the conjectural EP&MAG medium with local ME coupling, while the ICRs (44) describe the conjectural EP&MP medium with local ME coupling. Beyond a physical insight and necessary microscopic analysis at the present stage of study, let us suppose that we succeeded to realize local ME particles and create bianisotropic media with local properties based on a composition of such particles. If so, we might presume that we have a physical justification in using the above ICRs.

When a bianisotropic medium is time invariant and spatially homogeneous, the ICRs have a temporal and space convolution form. Based on the above consideration we introduce a notion of the local temporally dispersive bianisotropic media (LTDBM). These media are characterized by constitutive parameters satisfying the long-wavelength (quasistatic) limit. When this limit takes place, we can take advantage of the power-series expansion of the constitutive tensors over $\vec{k}$ in the region near $|\vec{k}| = 0$. The local temporally dispersive bianisotropic media are characterized by unique properties of balance of energy. These properties were shown in [47] based on an analysis of averaged quadratic forms in the Poynting theorem for the ME-coupling EP&MP medium.

For the quasimonochromatic fields $\vec{E} = \vec{E}_m(\tau) e^{i\omega t}$ and $\vec{H} = \vec{H}_m(\tau) e^{i\omega t}$, where complex amplitudes $\vec{E}_m(\tau)$ and $\vec{H}_m(\tau)$ are smooth functions of time, one obtains the standard-form energy balance equation

$$-\nabla \cdot \overline{\vec{S}} = \frac{\partial \overline{W}}{\partial \tau} + \overline{P} \tag{46}$$

for the LTDBM only when the following field-structure constraints for slowly time-varying amplitudes are imposed:

$$E^*_{m_i}(\tau) \frac{\partial H_{m_j}(\tau)}{\partial \tau} = H_{m_j}(\tau) \frac{\partial E^*_{m_i}(\tau)}{\partial \tau}. \tag{47}$$

In Eqs. (46), (47) $\overline{\vec{S}}$ is the average (on the period $2\pi/\omega$) Poynting vector, $\overline{W}$ is the average energy density, $\overline{P}$ is the average density of dissipation losses, subscripts $i$ and $j$ correspond the field components. When the above field-structure constrains are satisfied, one obtains the following average energy density in a LTDBM [47]:

$$\overline{W} = \frac{1}{4} \left\{ \frac{\partial \left(\omega \varepsilon^h_{ij}\right)}{\partial \omega} E^*_i E_j + \frac{\partial \left(\omega \mu^h_{ij}\right)}{\partial \omega} H^*_i H_j + \frac{\partial \left[\omega \left(\zeta^h_{ij} + \xi^h_{ij}\right)\right]}{\partial \omega} \left(H^*_i E_j\right)^h + \frac{\partial \left[\omega \left(\zeta^{ah}_{ij} - \xi^{ah}_{ij}\right)\right]}{\partial \omega} \left(H^*_i E_j\right)^{ah} \right\}, \tag{48}$$

where superscripts *h* and *ah* mean, respectively, "Hermitian" and "anti-Hermitian".

The field constrains should be imposed simultaneously to all field components. One can transform Eqs. (47) as follows:

$$\frac{E^*_{m_i}(\tau)}{H_{m_j}(\tau)} = \frac{\partial E^*_{m_i}(\tau)/\partial \tau}{\partial H_{m_j}(\tau)/\partial \tau} = \frac{dE^*_{m_i}(\tau)}{dH_{m_j}(\tau)}, \tag{49}$$

where $dE^*_{m_i}$ and $dH_{m_j}$ are differentials of the corresponding fields. Eqs. (49) show that there exist linear relations between the field amplitudes. In a general form, one can represent these relations as

$$E_{m_i}(\tau) = F_{ij} H^*_{m_j}(\tau), \tag{50}$$

where elements $F_{ij}$ of matrix $[F]$ are time-independent quantities. We will call matrix $[F]$ as a field-polarization matrix of a bianisotropic medium. This is an invariant (for certain average energy density). It is evident that components of matrix $[F]$ are complex quantities. In electromagnetic processes, electric and magnetic fields are mutually associated due to Maxwell's equations. At the same time, because of a presumed intrinsic ME effect in bianistropic media there exist additional (with respect to Maxwell's equations) associations between electric and magnetic fields described by certain wave-polarization parameters $F_{ij}$. To find parameters of polarization matrix $[F]$, one should solve a concrete electromagnetic problem based on the known medium constitutive parameters and boundary conditions. When the parameters of matrix $[F]$ are found, one can determine average energy density in a LTDBM [see Eq. (48)]. This is, in general, an integro-differential problem.

Let us consider the transversal field-polarization matrix $[F_\perp]$:

$$[F_\perp] \equiv \begin{bmatrix} \dfrac{E_{m_x}}{H^*_{m_x}} & \dfrac{E_{m_x}}{H^*_{m_y}} \\ \dfrac{E_{m_y}}{H^*_{m_x}} & \dfrac{E_{m_y}}{H^*_{m_y}} \end{bmatrix} = \begin{bmatrix} \dfrac{E_{m_x} H_{m_x}}{\left|H_{m_x}\right|^2} & \dfrac{E_{m_x} H_{m_y}}{\left|H_{m_y}\right|^2} \\ \dfrac{E_{m_y} H_{m_x}}{\left|H_{m_x}\right|^2} & \dfrac{E_{m_y} H_{m_y}}{\left|H_{m_y}\right|^2} \end{bmatrix}. \tag{51}$$

Evidently, in a general case this is neither Hermitian nor anti-Hermitian matrix.

The basic paradigm of the field-polarization effects in electromagnetics is elliptical polarization. Such polarization is obtained only for purely monochromatic radiation. If radiation is not purely monochromatic, one has polarization figures that are more complicated than ellipses. However, one can consider the case of elliptical polarization for quasimonochromatic waves when time of the averaging procedure $\tau$ essentially exceeds the RF period. We represent transverse fields of elliptically polarized waves as [48]:

$$\vec{E}_\perp = \left(\vec{p}^E - i\vec{q}^E\right)e^{i\omega t} \tag{52}$$

and

$$\vec{H}_\perp = \left(\vec{p}^H - i\vec{q}^H\right)e^{i\omega t}, \tag{53}$$

where $\vec{p}^E, \vec{q}^E, \vec{p}^H,$ and $\vec{q}^H$ are real vectors. We may write for components of matrix $[F_\perp]$:

$$F_{xx} = \frac{p_x^E p_x^H - q_x^E q_x^H - i\left(p_x^E q_x^H + p_x^H q_x^E\right)}{\left(p_x^H\right)^2 + \left(q_x^H\right)^2}, \tag{54}$$

$$F_{xy} = \frac{p_x^E p_y^H - q_x^E q_y^H - i\left(p_x^E q_y^H + p_y^H q_x^E\right)}{\left(p_y^H\right)^2 + \left(q_y^H\right)^2}, \tag{55}$$

$$F_{yx} = \frac{p_y^E p_x^H - q_y^E q_x^H - i\left(p_y^E q_x^H + p_x^H q_y^E\right)}{\left(p_x^H\right)^2 + \left(q_x^H\right)^2}, \qquad (56)$$

$$F_{yy} = \frac{p_y^E p_y^H - q_y^E q_y^H - i\left(p_y^E q_y^H + p_y^H q_y^E\right)}{\left(p_y^H\right)^2 + \left(q_y^H\right)^2}. \qquad (57)$$

Components of matrix $[F_\perp]$, being an invariant, describe complex scalar fields. There is a strong correlation between the amplitudes and phases of the electric and magnetic fields. The integral complex scalar field we will call the ME field. So one can presume that for electromagnetic waves propagating in a bianisotropic medium, there exists an invariant structure of the ME fields.

A generic solution of a system of equations $F_{ij} = 0$ may give polarization singularity. There is a system of eight nonlinear algebraic equations for eight unknown quantities $p_{x,y}^{E,H}$, $q_{x,y}^{E,H}$. The obtained solution will show the field polarization morphology for the ME field. For nonlinear algebraic equations, the generic conditions for existence of real solutions may be very complicated. A detailed mathematical analysis of a system of such nonlinear equations is beyond a scope of a present work and should be a subject of separate studies. A priori, one cannot be sure that there exists at least one real root. If there are no real solutions for a system of equations $F_{ij} = 0$, one can get a real root reducing an order of the equation by introduction of some additional conditions for unknown quantities. There can be, for example, certain conditions of excitation.

The above analysis of the field polarization properties arose from the field constrains (47) necessary to provide the energy balance relations in LTDBMs. In general, this gives the results different from the known studies of the electromagnetic-field polarization properties. In [49, 50], it was shown that for electric and magnetic fields in space there exist lines of singular polarization. In places of singularity one has or purely circular or purely linear polarizations. As it was found in [51], relativistically invariant line singularities of the full electromagnetic field are the Riemann-Silberstein vortices. In paper [52], devoted to an analysis of optical singularities in bianisotropic crystals, Berry chose $\vec{D}$ and $\vec{B}$ fields, rather than $\vec{E}$ and $\vec{H}$ fields. This makes an analysis simpler, since $\vec{D}$ and $\vec{B}$ are always transverse to the wave direction. It was shown that the singularities of $\vec{D}$ and $\vec{B}$ fields, being considered separately, are the circular polarization points. It follows from our analysis that singularity of the ME fields in bianisotropic media do not correspond to singularities considered in [49 – 52]. As one can see from Eqs. (54) – (57), possible field singularities in LTDMs are neither circular, nor linear polarization points. In singular points one will obtain special "polarization flowers" characterizing by strong correlations between parameters $p_{x,y}^E$ and $q_{x,y}^E$, from one side, and parameters $p_{x,y}^H$ and $q_{x,y}^H$, from the other side.

For complex-amplitude wave functions describing electromagnetic fields, the properties of helicity appear. This takes place due to superposition of not-equivalent purely left- and right-handed circularly polarized waves [50, 53]. The helicity can be defined as the sign of the projection of the angular momentum on the direction of the wave vector. It is evident that the ME field, describing by complex scalar fields, may also exhibit properties of helicity.

In a propagating-wave electromagnetic process, the transversal (with respect to the Poynting vector) electric and magnetic fields are in-phase at every time moment. The transversal-field energy density is the entire notion and there is no mutual transformation of the transversal-field electric and magnetic energies. On the other hand, magnetoelectricity may suppose coupling between the transversal-field energies of the electric and magnetic polarizations. As a result, for electromagnetic-wave processes in bianisotropic media one may have a mutual transformation (at time) of the transversal-field electric and magnetic energies.

In electromagnetics, a well-known impedance concept is regarded as a characteristic of the type of the fields. The main feature of the wave impedance, which arises immediately from Maxwell's equations, is its correlation with a power flow. Suppose now that the average Poynting vector is directed along $z$ axis (in a general case of a bianisotropic medium, the Poynting vector is not directed along the wave vector). We have

$$\bar{S}_z = \frac{1}{2}\text{Re}\,(\vec{E}_m \times \vec{H}_m^*)_z = \frac{1}{2}\text{Re}\,(E_{m_x} H_{m_y}^* - E_{m_y} H_{m_x}^*), \tag{59}$$

where $E_{m_x}, H_{m_y}, E_{m_y}$, and $H_{m_x}$ are the transversal field components with respect to the Poynting vector. Evidently, for propagating waves in a lossless medium one should have the following relations for the field amplitudes:

$$E_{m_x} = Z_0 H_{m_y} \quad \text{and} \quad E_{m_y} = -Z_0 H_{m_x}, \tag{60}$$

where $Z_0$ is a real-value characteristic impedance. $H_{m_y}$ has zero phase shift with respect to $E_{m_x}$. Also $H_{m_x}$ has zero phase shift with respect to $E_{m_y}$. Nevertheless, it is evident that in equations

$$E_{m_x} = F_{xy} H_{m_y}^* \quad \text{and} \quad E_{m_y} = F_{yx} H_{m_x}^*, \tag{61}$$

coefficients $F_{xy}$ and $F_{yx}$ are not real quantities in a general case of the field polarization.

Components of matrix $[F_\perp]$ are determined by a character of the two-dimensional transverse-field polarization. Taking into account Eqs. (60), one obtains

$$F_{xx} \equiv \frac{E_{m_x}}{H_{m_x}^*} = Z_0 \frac{H_{m_y}}{H_{m_x}^*} = -Z_0 \frac{E_{m_x}}{E_{m_y}^*}, \tag{62}$$

$$F_{xy} \equiv \frac{E_{m_x}}{H_{m_y}^*} = Z_0 \frac{H_{m_y}}{H_{m_y}^*} = Z_0 \frac{E_{m_x}}{E_{m_x}^*}, \tag{63}$$

$$F_{yx} \equiv \frac{E_{m_y}}{H_{m_x}^*} = -Z_0 \frac{H_{m_x}}{H_{m_x}^*} = -Z_0 \frac{E_{m_y}}{E_{m_y}^*}, \tag{64}$$

$$F_{yy} \equiv \frac{E_{m_y}}{H_{m_y}^*} = -Z_0 \frac{H_{m_x}}{H_{m_y}^*} = Z_0 \frac{E_{m_y}}{E_{m_x}^*}. \tag{65}$$

Based on Eqs. (62) – (65), it is possible to show that the following normalization condition takes place

$$|F_{xy}|^2 = |F_{yx}|^2 = -F_{xx}F_{yy}^* = -F_{xx}^*F_{yy} = Z_0^2. \tag{66}$$

We can represent now matrix $[F_\perp]$ as:

$$[F_\perp] = Z_0 \begin{bmatrix} \dfrac{H_{m_x} H_{m_y}}{|H_{m_x}|^2} & \dfrac{(H_{m_y})^2}{|H_{m_y}|^2} \\ -\dfrac{(H_{m_x})^2}{|H_{m_x}|^2} & -\dfrac{H_{m_x} H_{m_y}}{|H_{m_y}|^2} \end{bmatrix} = Z_0 \begin{bmatrix} -\dfrac{E_{m_x} E_{m_y}}{|E_{m_y}|^2} & \dfrac{(E_{m_x})^2}{|E_{m_x}|^2} \\ -\dfrac{(E_{m_y})^2}{|E_{m_y}|^2} & \dfrac{E_{m_x} E_{m_y}}{|E_{m_x}|^2} \end{bmatrix}. \tag{67}$$

The following relation arises immediately from Eq. (67):

$$\frac{F_{xx}}{F_{yy}} = -\frac{|E_{m_x}|^2}{|E_{m_y}|^2} = -\frac{|H_{m_y}|^2}{|H_{m_x}|^2}. \tag{68}$$

Since components of matrix $[F_\perp]$ are invariants, the above relation gives also an invariant quantity.

With help of characteristic impedance $Z_0$ we acquire a possibility to "untie" the electric and magnetic field polarizations in a ME complex scalar wavefunction. Based on Eqs. (67), we rewrite Eqs. (54) – (57) separately for the electric and magnetic field patterns of the ME field:

$$\frac{F_{xx}}{Z_0} = \frac{p_x^H p_y^H - q_x^H q_y^H - i(p_x^H q_y^H + p_y^H q_x^H)}{(p_x^H)^2 + (q_x^H)^2} = -\frac{p_x^E p_y^E - q_x^E q_y^E - i(p_x^E q_y^E + p_y^E q_x^E)}{(p_y^E)^2 + (q_y^E)^2}, \tag{69}$$

$$\frac{F_{xy}}{Z_0} = \frac{(p_y^H)^2 - (q_y^H)^2 - i2(p_y^H q_y^H)}{(p_y^H)^2 + (q_y^H)^2} = \frac{(p_x^E)^2 - (q_x^E)^2 - i2(p_x^E q_x^E)}{(p_x^E)^2 + (q_x^E)^2}, \tag{70}$$

$$\frac{F_{yx}}{Z_0} = -\frac{(p_x^H)^2 - (q_x^H)^2 - i2(p_x^H q_x^H)}{(p_x^H)^2 + (q_x^H)^2} = -\frac{(p_y^E)^2 - (q_y^E)^2 - i2(p_y^E q_y^E)}{(p_y^E)^2 + (q_y^E)^2}, \tag{71}$$

$$\frac{F_{yy}}{Z_0} = -\frac{p_x^H p_y^H - q_x^H q_y^H - i(p_x^H q_y^H + p_y^H q_x^H)}{(p_y^H)^2 + (q_y^H)^2} = \frac{p_x^E p_y^E - q_x^E q_y^E - i(p_x^E q_y^E + p_y^E q_x^E)}{(p_x^E)^2 + (q_x^E)^2}. \tag{72}$$

Evidently, in the direction along the Poynting vector, one has the same polarization ellipse morphology for the electric and magnetic fields. In this case, conditions $F_{ij} = 0$, determining polarization singularity, lead to overdetermined set of equations: one has a system of eight nonlinear algebraic equations for four unknown quantities $p_{x,y}^E$, $q_{x,y}^E$ (or $p_{x,y}^H$, $q_{x,y}^H$).

## 5. Microscopic and mesoscopic aspects of local ME particles

Till now, no necessary microscopic justifications of local ME particles – structural elements of BMMs – have been done. As we discussed above, formally, equations of macroscopic electrodynamics are equally applicable for media with electric polarization and magnetization – the EP&MAG media – as well as for media with electric polarization and magnetic polarization – the EP&MP media. At the same time, no ME couplings on the microscopic level are assumed in classical motion equations: no helical loops (recursion motions) are possible for bound electric charges and no interaction between linear electric and magnetic currents takes place.

The known propositions of bianisotropic microwave composites were made based on classical models which are rather contradictory. For example, very specific propositions of ME particles and chiral media made in [54] are based on hypothetical classical models: a model of a rotating electric dipole with oriented magnetic moments, a model of precessing magnetic moments with an attached electric dipole, a model of charged bead sliding freely on a spiral. The proposed models absolutely do not provide the reader with real physics of ME coupling between the microscopic electric and magnetic currents. The dielectric polarization is parity-odd and time-reversal-even. At the same time, the magnetization is parity-even and time-reversal-odd [7]. These symmetry relationships make questionable an idea of simple combination of two (electric and magnetic) dipoles to realize local ME particles for bianisotropic metamaterials.

One may suppose that the unified ME fields originated from a point ME particle (when such a particle is created) will not be the classical fields, but the quantum (quantum-like) fields. It means that the motion equations inside a local ME particle should be the quantum (quantum-like) motion equations with special symmetry properties. The fundamental discrete symmetries of parity (P), time reversal (T) and charge conjugation (C), and their violations in certain situations, are central in modern elementary particle physics, and in atomic and molecular physics. As a basic principle, the weak interaction is considered as the only fundamental interaction, which does not respect left-right symmetry. The mutual interaction of magnetic and electric charges in the dynamical construction of the elementary particles could lead in a natural way to the parity violation observed in weak interactions [16]. Atoms are chiral due to the parity-violating weak neutral current interaction between the nucleus and the electrons [55]. In crystalline solids, ME effect presumes symmetry breakdown in a structure of a medium. ME interactions with mutually perpendicular electric and magnetic dipoles in crystal structures arise from toroidal distributions of currents and are described by anapole moments [9]. It has been discussed that the reason why such anapole moments appear is Stone's spinning-solenoid Hamiltonian [56]. Magnetoelectric properties of Stone's Hamiltonian become apparent because of Berry's curvature of the electronic wavefunctions. In recent theories of spin waves in magnetic-order crystals, a Berry curvature is stated as playing a key role [57, 58]. The Berry phase may also influence the properties of magnons. If the magnetic medium in which the magnon is propagating is spatially non uniform, a Berry phase may be accumulated along a closed circuit in space. It has been recently indicated [59] that the geometric Berry phase due to a non-coplanar texture of the magnetization of a ferromagnetic ring would affect the dispersion of magnons, lifting the degeneracy of clockwise and anticlockwise propagating magnons. It was found [59] that the magnetization transport by magnons in a noncollinear spin structure is accompanied by an electric polarization. This electric polarization can be experimentally observed not only in the vicinity of the mesoscopic ring [59] but also in the vicinity of the magnetic wire [60]. Moving magnetic dipoles represent an electric dipole moment [61] and are therefore affected by electric fields. Such a ME effect in magnetic nanostructures is, in fact, the relativistic effect of a transformation of magnetization to the electric field in the moving frame. At the same time, because of the electronic structure of the material, the magnitude of such ME coupling can be much larger than a bare relativistic effect [62].

The above microscopic and mesoscopic aspects of chirality and magnetoelectricity should, certainly, be related to the problems of local ME particles and the unified ME fields originated from such point sources, and necessarily should become the main subject for realization of BMMs. In Introduction, we pointed out that the concept of a bianisotropics, coined in 1968 by Cheng and Kong [1], was aimed, in particular, to unify two separate branches of research on moving media and ME crystals. One may paraphrase this concept of bianisotropics in other words: possible unification of electromagnetic processes of dipole motions and symmetry breaking phenomena. In such a sense, following the results of recent studies, we can state now that spectral properties of magnetic-dipolar modes (MDM) in ferrite disks may put us into proper way. It was shown that magnetic dipole motion processes in a normally magnetized ferrite disk are characterized by handedness properties. Both aspects – the anapole-moment properties of ME crystals and a ME effect in magnetic nanostructures – become evident for magnetic-dipolar oscillations in ferrite disks.

In microwaves, ferrite resonators with multi-resonance MDM [or magnetostatic (MS)] oscillations may have sizes two-four orders less than the free-space EM wavelength at the same frequency. In quasi-two-dimensional systems, the dipolar interaction can play an essential role in determine the magnetic properties. MS ferromagnetism has a character essentially different from exchange ferromagnetism [34]. This statement finds strong confirmation in confinement phenomena of magnetic-dipolar oscillations. The dipole interaction provides us with a long-range mechanism of interaction, where a magnetic medium is considered as a continuum. Contrary to an exchange spin wave, in magnetic-dipolar waves the local fluctuation of magnetization does not propagate due to interaction between the neighboring spins. There should be certain propagating fields – the MS fields – which cause and govern propagation of magnetization fluctuations. In other words, space-time magnetization fluctuations are corollary of the propagating MS fields, but there are no magnetization waves. The boundary conditions should be imposed on the MS field and not on the RF magnetization. When field differences across the sample become comparable to the bulk demagnetizing fields the local-oscillator approximation is no longer valid, and indeed under certain circumstances, entirely new spin dynamics behavior can be observed.

The character of the experimental multi-resonance absorption spectra for small ferrite disks (obtained with respect to a DC bias magnetic field $H_0$) [63] leads to a clear conclusion that the energy of a source of a DC magnetic field is absorbing "by portions" or discretely, in other words. There are the regular spectra with properties similar to an atomic-like $\delta$- function density of states. The theory of the magnetic-dipolar mode spectra is based on the notion of MS-potential wave functions $\psi$. MS-potential wave functions may acquire a special physical meaning in a case of MDM oscillations in a normally magnetized ferrite disk. As it has been shown recently [64], in such a sample the confinement effect leads to the quantum properties described by the Schrödinger-like equation for a MS-potential wave function. A ferrite disk with the MDM oscillating spectra is a mesoscopic system in a sense that such a system is sufficiently big compared to atomic and lattice scales, but sufficiently small that quantum mechanical phase coherence is preserved around the whole sample. For a case of a MDM ferrite disk one has the quantized-like oscillating system which preserves the coherence. Following an idea that external interactions should increase the probability of transition of a quantum system to other states leading to line broadening, it was experimentally shown in [65] how the environment may cause decoherence for magnetic oscillations.

Quantum effects are rarely observed through macroscopic measurements because statistical averaging over many states usually masks all evidence of discreteness. Notable exceptions include the dc and ac Josephson effects, the dc Haas-van Alphen effect and the quantum Hall effect. There are certain aspects that allow distinguishing the quantum-confinement oscillations from the classical-confinement oscillations. In a case of the quantum confinement one has: (a) discrete portions of energy supplying by an external source, (b) the energy-egenstate spectral problem (discrete energy eigenstates, the Hilbert functional space), (c) complete-set scalar wave functions allowing to introduce the notion of probability, and, as a result, (d) all physical quantities (observables) are related to proper differential operators. Recent studies show a macroscopic effect of quantum coherence for MDM oscillations in normally magnetized thin-film ferrite disks.

In recent experiments it was shown that MDM oscillations in a normally magnetized ferrite disk are strongly affected by a normal component of the external RF electric field [65]. The observed multi-resonance absorption spectra are due to the eigen-electric-moment oscillations caused by the motion processes in a ferrite resonator. Since the RF electric field does not change sign under time inversion, the eigen electric moment should also be characterized by the time-reversal-even properties. As it was theoretically predicted [66, 67], MDM oscillations in a ferrite disk are accompanied with specific surface magnetic currents, which should cause the parity-violating perturbation and, as a result, should lead to appearance of the eigen electric moments in disk ferrite particles. One has special symmetry properties of the anapole moment [polar (electric) symmetry, i.e. the parity-odd, and time-reversal-even symmetry]. Surface magnetic currents (resulting in the anapole moments) in a ferrite disk appear due to symmetry breaking for magnetic-dipolar oscillating modes [66, 67].

Special symmetry properties of MDMs become evident when one considers the MS-wave propagation in a helical coordinate system. For solutions of the Landau-Lifshitz (LL) equation, one has different signs of the off-diagonal components of the permeability tensor in the right-handed and left-handed helical coordinate systems. Lifting the degeneracy of clockwise and anticlockwise propagating MS modes becomes apparent when one considers the LL equation for harmonic RF magnetization: $i\omega \vec{m} + \gamma \vec{m} \times \vec{H}_0 = -\gamma \vec{M}_0 \times \vec{H}$ with respect to a helical coordinate system. For given directions of bias magnetic field $\vec{H}_0$ and saturation magnetization $\vec{M}_0$, and for given RF magnetic field $\vec{H}$ and RF magnetization $\vec{m}$, one has opposite signs for vector products with respect to the right-handed and left-handed helical coordinate systems. This shows that for solutions of the LL equation: $\vec{m} = \ddot{\chi} \cdot \vec{H}$ there are different signs of off-diagonal components of the susceptibility tensor $\ddot{\chi}$ with respect to the right-handed and the left-handed helical coordinate systems. In the Waldron's helical coordinate system $(r, \phi, \zeta)$ [68], one has four types of helical waves (the forward and backward waves with right-handed and left-handed rotations) propagating in a ferrite rod [69]:

$$\begin{aligned} \psi^{(1)} &\sim e^{-iw\phi} e^{-i\beta\zeta}, \\ \psi^{(2)} &\sim e^{+iw\phi} e^{-i\beta\zeta}, \\ \psi^{(3)} &\sim e^{+iw\phi} e^{+i\beta\zeta}, \\ \psi^{(4)} &\sim e^{-iw\phi} e^{+i\beta\zeta}, \end{aligned} \tag{73}$$

where $w$ and $\beta$ are wavenumbers corresponding to $\phi$ and $\zeta$ coordinates. For a smooth infinite ferrite rod, the four-type helical waves are basic solutions which cannot be mutually transformed. Situation, however, becomes different in a case of a ferrite disk where the quantity of pitch can be determined by the virtual "reflection" planes [69, 70]. For different types of external RF fields acting on a ferrite disk, different cases of interactions between the helical waves are possible. In particular, one may have two resonances. The first resonance may appear due to $\psi^{(1)} \leftrightarrow \psi^{(4)}$ interaction. The second resonance – due to $\psi^{(2)} \leftrightarrow \psi^{(3)}$ interaction. These resonances, being characterized by different directions of azimuth rotation, are conventionally called as the "left" resonance and the "right" resonance. The obtained oscillations of helical MS-wave harmonics in a ferrite disk should be considered as adiabatic variations with respect to a dynamical precession process described by the LL equation. One has angle change in the transverse component of a magnetic moment when the field about which it precesses is varied adiabatically through a closed loop. Because of these adiabatic variations, MS-potential eigenfunctions should, in general, pick up the Berry phase. Fig. 3 illustrates the RF magnetization evolution for the "left" ($\psi^{(1)} \leftrightarrow \psi^{(4)}$ interaction) resonance [70].

The properties of the "left" and "right" resonances become evident from explicit analysis in a cylindrical coordinate system. The confinement effect for magnetic-dipolar oscillations requires proper phase relationships to guarantee single-valuedness of the wave functions. To compensate for sign ambiguities and thus to make wave functions single valued we should add a vector-potential-type term to the MS-potential Hamiltonian. The corresponding flux of pseudo-electric field $\vec{\in}$ (the gauge field) through a circle of disk radius $\Re$ is [67]

$$\int_S \vec{\in} \cdot d\vec{S} = \oint_{C=2\pi\Re} \vec{A}_\theta^m \cdot d\vec{C} = \Phi^e, \tag{74}$$

where $\Phi^e$ is the flux of the pseudo-electric field. The energy levels are periodic in the electric flux $\Phi^e$. There should be the positive and negative fluxes. These different-sign fluxes should be inequivalent to avoid the cancellation. The transformation restores the single valuedness, but now there is a nonzero vector-potential-type term $A_\theta^m$. The value $\oint_C \vec{A}_\theta^m \cdot d\vec{C} \neq 0$ can be observable. An analysis of a phase factor that makes the states single valued and so makes a total Hamiltonian to be conserved is related to a topological effect in a closed system. In this case the results are gauge invariant and the Stokes theorem can be used.

In such a closed system, there should be a certain internal mechanism which creates a non-zero vector potential $\vec{A}_\theta^m$. This internal mechanism appears due to a singular annual magnetic field, $\vec{H}_\theta$, and the off-diagonal component of the permeability tensor, $\mu_a$, resulting in an effective circular magnetic current, $i^m$. Circulation of current $i^m$ along contour $C$ may give a nonzero quantity. The fields existing inside a ferrite disk form a very special field structure outside a disk. Because of nonzero circulation of current $i^m$, one has an electric moment $\vec{a}^e$ of a whole ferrite disk resonator. An electric moment $\vec{a}^e$ is the parity-odd time-reversal-even function [65 - 67]. There are the symmetry properties of the anapole moment [10].

Together with the above case of $\psi^{(1)} \leftrightarrow \psi^{(4)}, \psi^{(2)} \leftrightarrow \psi^{(3)}$ resonances other types of interactions of helical modes may take place. In an external RF magnetic field one may have fractures of helical waves on "reflection" points on disk surfaces leading to appearance of surface magnetic dipoles. An azimuth rotation of these surface dipoles leads to appearance of electric dipoles [59, 61]. This may explain the observed ME effect in ferrite disks with surface electrodes [71]. This microwave effect of local ME coupling was theoretically predicted in [72]. One can see that in a case of a ferrite disk + wire particle [71] there are the ME modes characterizing by oscillations in metal-wire and ferrite-disk subsystems. An experimental object (a ferrite disk + wire particle) can be modeled as a triple of vectors: an axial vector of a magnetic bias field $\vec{H}_0$, a polar vector of an electric moment $\vec{p}^e$, and an axial vector of a magnetic moment $\vec{p}^m$. Evidently, this triple of vectors is not invariant under the classical PT transformation but is invariant under the non-classical CPT transformation (Fig. 4).

## 6. Conclusion

Following our classification, the BMMs appear as special-type MEMs. A combination of two physical notions – near-field manipulation and chirality – should underlie the main concept of BMMs. From classical laws these two notions are in an evident contradiction. The known manifestation of chirality in classical electromagnetics is due to effects of the field non-locality. When, however, one formally supposes existence of local bianisotropic media, he finds unique properties of the field polarization structure.

An assumption of the local cross-polarization effect will inevitably lead to a special ME field surrounding a ME particle. This may follow from a simple consideration of free-space Maxwell equations when external sources are taken to be external polarization $\vec{P}(\vec{r},t)$ and external magnetization $\vec{M}(\vec{r},t)$. If for monochromatic waves we assume external probes as local electric and magnetic dipoles:

$$\vec{P}(\vec{r},\omega) = \vec{\mathrm{p}}^e(\omega)\, \delta(\vec{r}-\vec{r}_0), \tag{75}$$

$$\vec{M}(\vec{r},\omega) = \vec{\mathrm{p}}^m(\omega)\, \delta(\vec{r}-\vec{r}_0), \tag{76}$$

solutions of Maxwell equations are (see e.g. [73])

$$\vec{E}(\vec{r},\omega) = \left[k_0^2 \vec{\mathrm{p}}^e + \vec{\nabla}\left(\vec{\mathrm{p}}^e \cdot \vec{\nabla}\right) + ik_0 \nabla \times \vec{\mathrm{p}}^m\right] \frac{e^{ik_0|\vec{r}-\vec{r}_0|}}{|\vec{r}-\vec{r}_0|}, \tag{77}$$

$$\vec{H}(\vec{r},\omega) = \left[k_0^2 \vec{\mathrm{p}}^m + \vec{\nabla}\left(\vec{\mathrm{p}}^m \cdot \vec{\nabla}\right) - ik_0 \nabla \times \vec{\mathrm{p}}^e\right] \frac{e^{ik_0|\vec{r}-\vec{r}_0|}}{|\vec{r}-\vec{r}_0|}, \tag{78}$$

where $k_0 = \omega/c$. In these equations, no coupling between local dipoles $\vec{\mathrm{p}}^e$ and $\vec{\mathrm{p}}^m$ is presupposed. So no ME coupling between electric and magnetic fields can be presupposed as well. When one accepts experimental data of local ME coupling in ferrite disks with surface electrodes shown in [71], he certainly should take into consideration the symmetry properties analyzed in Fig.4. This evidently results in non-classical symmetry properties of local ME fields.

In finite temporally dispersive media, quasistatic oscillations with broken symmetry between the electric and magnetic fields may take place. The basic physical characteristics of ferrite ME particles arise from MDM macroscopic quantum (quantum-like) phenomena with special topological effects. External adiabatic parameters appear due to external RF fields. In known microwave experiments, ferrite ME particles were placed in cavity fields. The staircase demagnetization energy in MDM ferrite disks, observed in experiments [63, 65, 71], may appear as coherent emission of microwave power forming geometry dependent standing waves in the cavity. To analyze interaction of ferrite ME particles with electromagnetic fields one should use methods of quantum electrodynamics. At present, the methods of cavity QED for ferrite ME particles have not been developed. This should be a subject for future special studies. When a local ME particle (based on MDM-oscillation spectra) is placed inside a cavity, quantization of electromagnetic fields takes place. Since the near field of a ME particle has symmetry breaking, the mapping of the ME-particle field into the cavity EM field will lead to symmetry breaking of the last one. So the cavity field may acquire the properties of helicity. There should be a certain correlation between symmetry properties of intrinsic motion processes in a local ME particle and symmetry properties of the ME near fields.

The problem of BMMs is, in fact, the problem of unification of electromagnetism and magnetoelectricity. Comprehensive solution of such a problem should be found through understanding a fundamental mechanism of a "junction" of electricity and magnetism in a point source – a local ME particle.

**References**

[1] D.K. Cheng and J.A. Kong, Proc. IEEE **56**, 248 (1968); J. Appl. Phys. **39**, 5792 (1968).
[2] *Advances in Complex Materials*, Eds. A Priou, A. Sihvola, S. Tretyakov, and A. Vinogradov (Dordrecht, Boston, London: Kluwer Academic Publ., 1997).
[3] A. Ishimaru *et al.*, IEEE Trans. Antennas Propag. **51**, 2550 (2003).
[4] R.E. Collin, *Field Theory of Guided Waves* (IEEE, New York, 1991).
[5] R.E. Raab and O.L. De Lange, *Multipole Theory in Electromagnetism* (Oxford, 2005).
[6] E.O. Kamenetskii, Phys. Rev. E **58**, 7965 (1998).
[7] J.D. Jackson, *Classical Electrodynamics*, 2nd ed. (Wiley, New York, 1975).
[8] A.P. Cracknell, *Magnetism in Crystalline Materials* (Oxford, Pergamon, 1975).
[9] P. Carra, J. Phys. A: Math. Gen. **37**, L183 (2004).
[10] Ya. B. Zel'dovich, Zh. Eksp. Teor. Fiz. **33**,1531 (1957) [Sov. Phys. JETP **6**, 1184 (1958)].
[11] W. C. Naxton, C.- P. Liu, and M. J. Rumsey-Musolf, Phys. Rev. C **65**, 045502 (2002).
[12] M. A. Bouchiat and C. Bouchiat, Eur. Phys. J. D **15**, 5 (2001).
[13] R. R. Lewis, Phys. Rev. A **49**, 3376 (1994).
[14] M.J. Padgett and L. Allen, Opt. Quant. Electron. **31**, 1 (1999).
[15] J. Salo *et al.*, J. Opt. A: Pure Appl. Opt. **4**, S161 (2002).
[16] D. Fryberger, Foundation of Physics **19**, 125 (1989).
[17] L.D. Landau and E.M. Lifshitz, *Electrodynamics of Continuous Media*, 2$^{nd}$ ed. (Pergamon, Oxford, 1984).
[18] J.B. Pendry *et al.*, IEEE Trans. Microw. Theory Techn. **47**, 2075 (1999).
[19] J.B. Pendry *et al.*, Phys. Rev. Lett. **76**, 4773 (1996).
[20] J.B. Pendry *et al.*, J. Phys. : Condens. Matter **10**, 4785 (1998).
[21] T. Kempa *et al.*, J. Appl. Phys. **98**, 034310 (2005).
[22] S. Riikonen *et al.*, Phys. Rev. B **71**, 235104 (2005).
[23] D,R, Smith *et al.*, Phys. Rev. Lett. **84**, 4184 (2000).
[24] G. Markovich *et al.*, Acc. Chem. Res. **32**, 415 (1999).
[25] J.B. Pendry, Phys. Rev. Lett. **85**, 3966 (2000); Optics & Photonic News, September 2004, pp. 33-37.
[26] C. Girard *et al.*, Rep. Prog. Phys. **63**, 893 (2000).
[27] C. Girard and A. Dereux, Rep. Prog. Phys. **59**, 657 (1996).
[28] F.J. Garcia de Abajo, Phys. Rev. B **60**, 6086 (1999).
[29] D. Maystre and S. Enoch, J. Opt. Soc. Am. A **21**, 122 (2004).
[30] S. Tretyakov *et al.*, Photonics Nanostructures – Fundament. Appl. **3**, 107 (2005).
[31] B.T. Rosner and D.W. van der Weide, Rev. Sci. Instr. **73**, 2505 (2002).
[32] J.Q. Lu and A.A. Maradudin, Phys Rev. B **42**, 11159 (1990).
[33] D.R. Fredkin and I.D. Mayergoyz, Phys. Rev. Lett. **91**, 253902 (2003).
[34] A. Gurevich and G. Melkov, *Magnetic Oscillations and Waves* (CRC Press, New York, 1996).
[35] F. Olyslager and I.V. Lindell, IEEE Antennas Propag. Magazine, **44**, 48 (2002).
[36] E.J. Post, *Formal Structure of Electromagnetics* (North-Holland, Amsterdam, 1962).
[37] F. Gronwald *et al.*, arXiv: physics/0506219 (2005).
[38] M.I. Ryazanov, *Condensed Matter Electrodynamics* (Nauka, Moscow, 1984) (In Russian).

[39] P. Politi and M.G. Pini, Phys. Rev. B **66**, 214414 (2002).
[40] K. Rivkin *et al*., Phys. Rev. B **70**, 184410 (2004).
[41] V.G. Levich, *Course of Theoretical Physics*, *Vol. 1* (Nauka, Moscow, 1969) (In Russian).
[42] T.G. Mackay *et al*., Phys. Rev. E **62**, 6052 (2000).
[43] R. Marques *et al*., Phys. Rev. B **65**, 144440 (2002).
[44] N. Katsarakis et al., Appl. Phys. Lett. **84**, 2943 (2004).
[45] X. Chen *et al*., Phys. Rev. E **71**, 046610 (2005).
[46] V.M. Agranovich and V.L. Ginzburg, *Crystal Optics with Spatial Dispersion and Excitons*, 2$^{nd}$ ed, (Springer-Verlag, New York, 1984).
[47] E.O. Kamenetskii, Phys. Rev. E **54**, 4359 (1996).
[48] M. Born and E.W. Wolf, *Principles of Optics* (Pergamon, Oxford, 1959)
[49] J.F. Nye, Proc. R. Soc. Lond. A **389**, 279 (1983).
[50] J.F. Nye and J.V. Hajnal, Proc. R. Soc. Lond. A **409**, 21 (1987).
[51] I. Bialinicky-Birula and Z. Bialinicky-Birula, Phys. Rev. A **67**, 062114 (2003).
[52] M.V. Berry, Proc. R. Soc. Lond. A **461**, 2071 (2005).
[53] G. Kaiser, J. Opt. A: Pure Appl. Opt. **6**, S243 (2004).
[54] J. Baker-Jarvis and P. Kabos, Phys. Rev. E **64**, 056127 (2003).
[55] R. A. Hergstrom et al., Am. J. Phys. **56**, 1086 (1988).
[56] M. Stone, Phys. Rev. D **33**, 1191 (1986).
[57] Q. Niu and L. Kleinman, Phys. Rev. Lett. **80**, 2205 (1998).
[58] R. Gebauer and S. Baroni, Phys. Rev. B **61**, R6459 (2000).
[59] F. Schütz, M. Kollar, and P. Kopietz, Phys. Rev. Lett. **91**, 017205 (2003).
[60] F. Meier and D. Loss, Phys. Rev. Lett. **90**, 167204 (2003).
[61] J.E. Hirsch, Phys. Rev. B **60**, 14787 (1999).
[62] P. Bruno and V.K. Dugaev, Phys. Rev. B **72**, 241302(R) (2005).
[63] J.F. Dillon Jr., J. Appl. Phys. **31**, 1605 (1960); T. Yukawa and K. Abe, J. Appl. Phys. **45**, 3146 (1974).
[64] E.O. Kamenetskii, Phys. Rev. E **63**, 066612 (2001); E.O. Kamenetskii, R. Shavit, and M. Sigalov, Europhys. Lett. **64**, 730 (2003); E.O. Kamenetskii, M. Sigalov, and R. Shavit, J. Phys.: Condens. Matter **17**, 2211 (2005).
[65] E.O. Kamenetskii, A.K. Saha, and I. Awai, Phys. Lett. A **332**, 303 (2004).
[66] E.O. Kamenetskii, Europhys. Lett. **65**, 269 (2004).
[67] E.O. Kamenetskii, Phys. Rev. E **73**, 016602 (2006).
[68] R.A. Waldron, Quarterly J. Mech. Appl. Math. **11**, 438 (1958).
[69] E.O. Kamenetskii, Physica E **29**, 350 (2005); E.O. Kamenetskii, J. Magn. Magn. Mater. **302**, 137 (2006).
[70] E.O. Kamenetskii, cond-mat/0505717.
[71] E. O. Kamenetskii, I. Awai, and A.K. Saha, Microw. Opt. Technol. Lett., **24**, 56 (2000); A.K. Saha, E.O. Kamenetskii, and I. Awai, Phys. Rev. E **64**, 056611 (2001); A.K. Saha, E.O. Kamenetskii, and I. Awai, J. Phys. D: Appl. Phys. **35**, 2484 (2002).
[72] E. O. Kamenetskii, Microw. Opt. Technol. Lett. **11**, 103 (1996); E. O. Kamenetskii, Phys. Rev. E **57,** 3563 (1998).
[73] G.S. Agarwal, Phys. Rev. A **11**, 230 (1974).

**Figure captions**

Fig. 1. Model of an EP&MP material

Fig. 2. Model of a questionable EP&MP medium with local ME coupling

Fig. 3. Illustration of RF magnetization evolution for the "left" ($\psi^{(1)} \leftrightarrow \psi^{(4)}$ interaction) resonance

Fig. 4. The CPT invariance of a system of two axial and one polar vectors

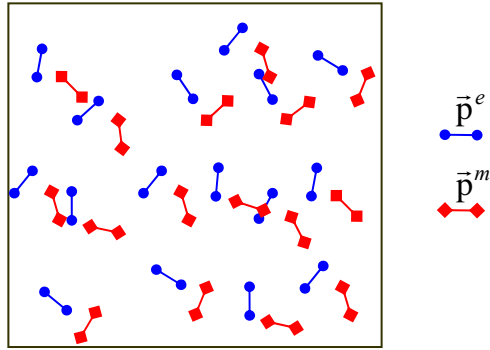

Fig. 1. Model of an EP&MP material

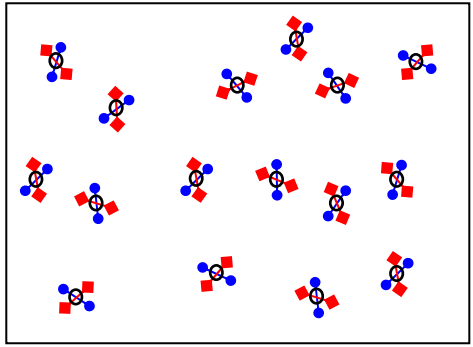

Fig. 2. Model of a questionable EP&MP medium with local ME coupling

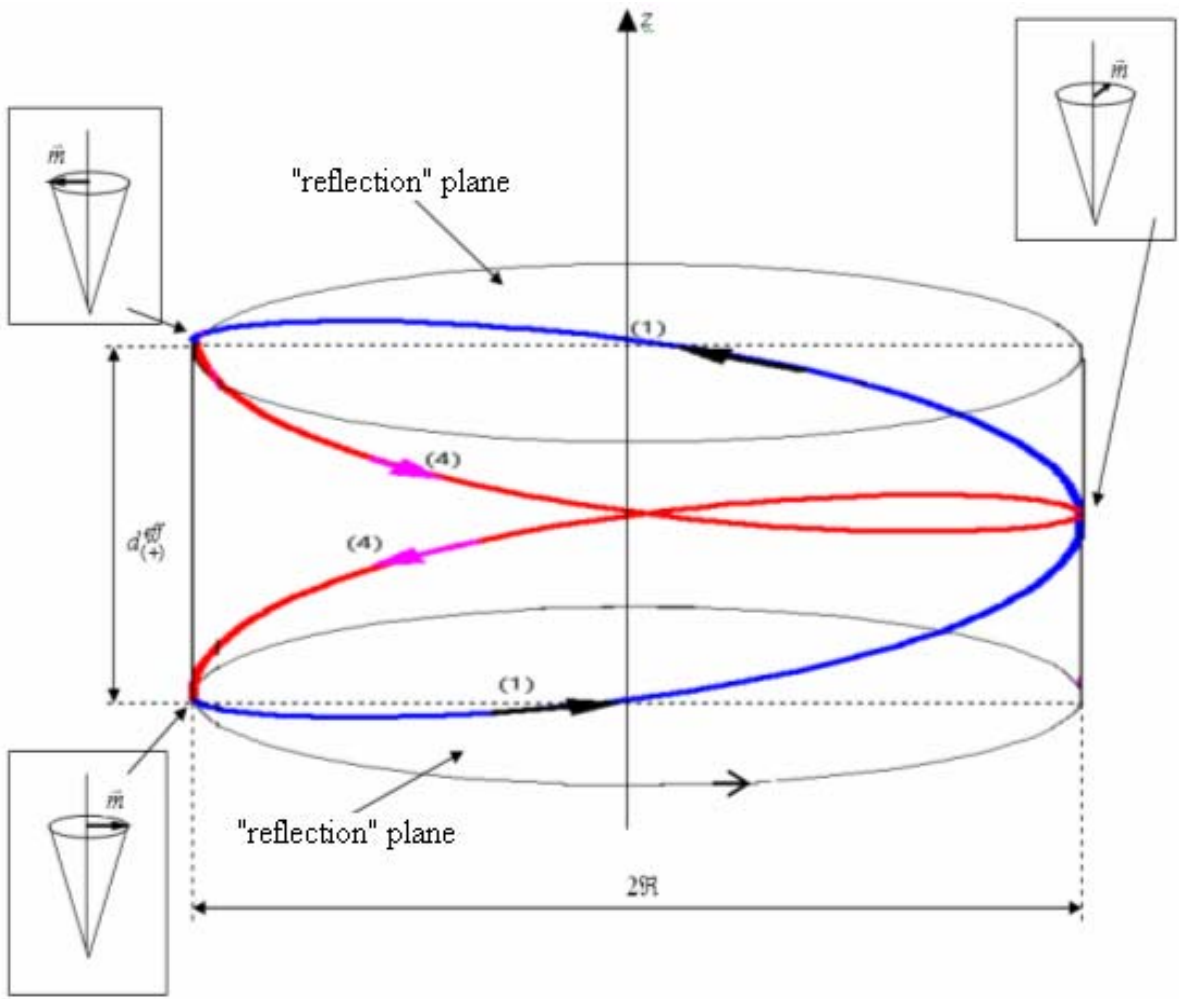

Fig. 3. Illustration of RF magnetization evolution for the "left" ($\psi^{(1)} \leftrightarrow \psi^{(4)}$ interaction) resonance

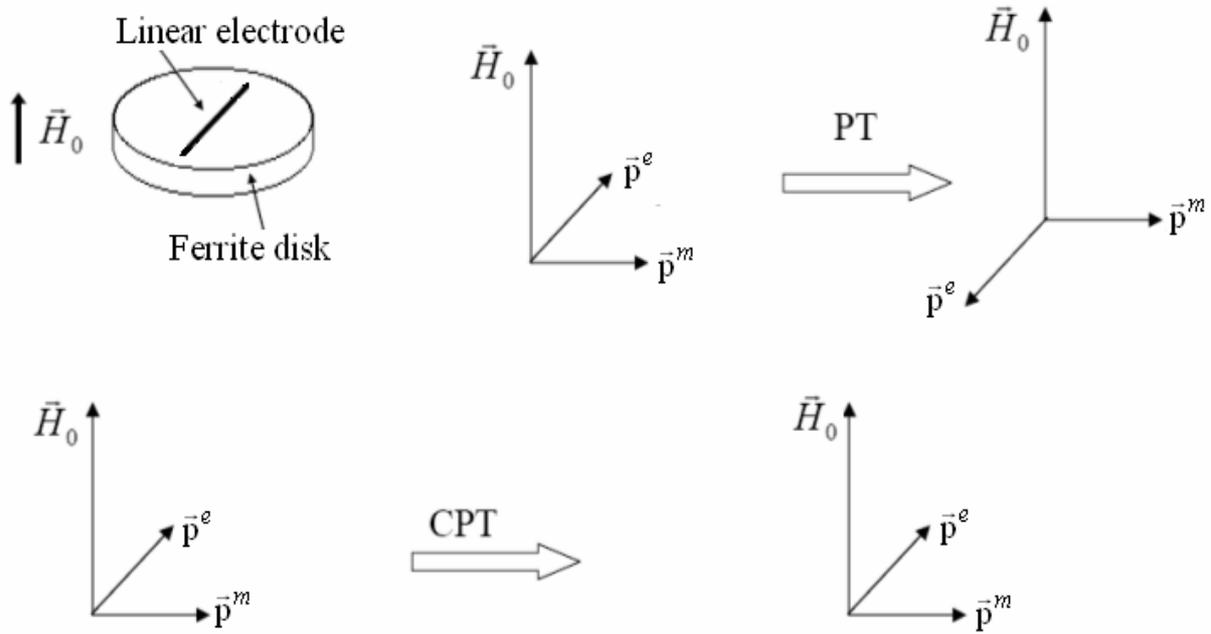

Fig. 4. The CPT invariance of a system of two axial and one polar vectors